\documentclass[twocolumn, superscriptaddress, amsmath, amssymb, aps, prapplied, floatfix,longbibliography]{revtex4-2}

\usepackage{graphicx}
\usepackage{dcolumn}
\usepackage{bm}
\usepackage{siunitx}
\usepackage{textcomp}
\usepackage{xcolor}
\usepackage{float}
\usepackage{array}
\newcommand{\ket}[1]{\left| #1 \right>} % for Dirac kets
\newcolumntype{L}[1]{>{\raggedright\let\newline\\\arraybackslash\hspace{0pt}}m{#1}}
\newcolumntype{C}[1]{>{\centering\let\newline\\\arraybackslash\hspace{0pt}}m{#1}}
\newcolumntype{R}[1]{>{\raggedleft\let\newline\\\arraybackslash\hspace{0pt}}m{#1}}

\newcommand{\titlename}{Development of a Boston-area 50-km fiber quantum network testbed}

\begin{document}

\title{\titlename}

\author{Eric Bersin}
\email{Eric.Bersin@ll.mit.edu}
\affiliation{Lincoln Laboratory, Massachusetts Institute of Technology, Lexington, MA 02421, USA}
\affiliation{Research Laboratory of Electronics, Massachusetts Institute of Technology, Cambridge, MA 02139, USA}

\author{Matthew Grein}
\affiliation{Lincoln Laboratory, Massachusetts Institute of Technology, Lexington, MA 02421, USA}
 
\author{Madison Sutula}
\affiliation{Department of Physics, Harvard University, Cambridge, MA 02138, USA}
 
\author{Ryan Murphy}
\affiliation{Lincoln Laboratory, Massachusetts Institute of Technology, Lexington, MA 02421, USA}
 
\author{Yan Qi Huan}
\affiliation{Department of Physics, Harvard University, Cambridge, MA 02138, USA}
 
\author{Mark Stevens}
\affiliation{Lincoln Laboratory, Massachusetts Institute of Technology, Lexington, MA 02421, USA}

\author{Aziza Suleymanzade}
\affiliation{Department of Physics, Harvard University, Cambridge, MA 02138, USA}
 
\author{Catherine Lee}
\affiliation{Lincoln Laboratory, Massachusetts Institute of Technology, Lexington, MA 02421, USA}

\author{Ralf Riedinger}
\affiliation{Department of Physics, Harvard University, Cambridge, MA 02138, USA}
\affiliation{Institut f\"{u}r Laserphysik und Zentrum f\"{u}r Optische Quantentechnologien, Universit\"{a}t Hamburg, 22761 Hamburg, Germany}
\affiliation{The Hamburg Centre for Ultrafast Imaging, 22761 Hamburg, Germany}

\author{David J.~Starling}
\affiliation{Lincoln Laboratory, Massachusetts Institute of Technology, Lexington, MA 02421, USA}

\author{Pieter-Jan Stas}
\affiliation{Department of Physics, Harvard University, Cambridge, MA 02138, USA}

\author{Can M. Knaut}
\affiliation{Department of Physics, Harvard University, Cambridge, MA 02138, USA}
 
\author{Neil Sinclair}
\affiliation{John A. Paulson School of Engineering and Applied Sciences, Harvard University, Cambridge, MA 02138, USA}

\author{Daniel R. Assumpcao}
\affiliation{John A. Paulson School of Engineering and Applied Sciences, Harvard University, Cambridge, MA 02138, USA}

\author{Yan-Cheng Wei}
\affiliation{Department of Physics, Harvard University, Cambridge, MA 02138, USA}

\author{Erik N. Knall}
\affiliation{John A. Paulson School of Engineering and Applied Sciences, Harvard University, Cambridge, MA 02138, USA}

\author{Bartholomeus Machielse}
\affiliation{Department of Physics, Harvard University, Cambridge, MA 02138, USA}
\affiliation{AWS Center for Quantum Networking, Boston, MA 02135, USA}

\author{Denis D. Sukachev}
\affiliation{Department of Physics, Harvard University, Cambridge, MA 02138, USA}
\affiliation{AWS Center for Quantum Networking, Boston, MA 02135, USA}

\author{David S. Levonian}
\affiliation{Department of Physics, Harvard University, Cambridge, MA 02138, USA}
\affiliation{AWS Center for Quantum Networking, Boston, MA 02135, USA}

\author{Mihir K. Bhaskar}
\affiliation{Department of Physics, Harvard University, Cambridge, MA 02138, USA}
\affiliation{AWS Center for Quantum Networking, Boston, MA 02135, USA}

\author{Marko Lon\v{c}ar}
\affiliation{John A. Paulson School of Engineering and Applied Sciences, Harvard University, Cambridge, MA 02138, USA}

\author{Scott Hamilton}
\affiliation{Lincoln Laboratory, Massachusetts Institute of Technology, Lexington, MA 02421, USA}

\author{Mikhail Lukin}
\affiliation{Department of Physics, Harvard University, Cambridge, MA 02138, USA}
 
\author{Dirk Englund}
\affiliation{Research Laboratory of Electronics, Massachusetts Institute of Technology, Cambridge, MA 02139, USA}
 
\author{P.~Benjamin Dixon}
\affiliation{Lincoln Laboratory, Massachusetts Institute of Technology, Lexington, MA 02421, USA}

\begin{abstract}
Distributing quantum information between remote systems
will necessitate the integration of emerging quantum components with existing communication infrastructure. This requires understanding the channel-induced degradations of the transmitted quantum signals, beyond the typical characterization methods for classical communication systems.
Here we report on a comprehensive characterization of a Boston-Area Quantum Network \mbox{(BARQNET)} telecom fiber testbed, measuring the time-of-flight, polarization, and phase noise imparted on transmitted signals. 
We further design and demonstrate a compensation system that is both resilient to these noise sources and compatible with integration of emerging quantum memory components on the deployed link.
These results have utility for future work on the BARQNET as well as other quantum network testbeds in development, enabling near-term quantum networking demonstrations and informing what areas of technology development will be most impactful in advancing future system capabilities.
\end{abstract}

\maketitle

\section{Introduction}
The continued development of networked quantum systems~\cite{Awschalom2021} will support applications in distributed quantum processing~\cite{Monroe2014, Nickerson2014, Litinski2022}, enhanced sensing~\cite{Gottesman2012, Khabiboulline2019, Komar2014, Nichol2022, Haldar2023, Zhang2021}, and secure communications~\cite{Muller_1993,Pirandola_2020,Xu2020, Skoric2017}. Connecting these systems across distances ranging from tens of meters to hundreds of kilometers requires the careful testing and integration of a range of quantum and classical technologies.  
Key challenges are quantifying and overcoming the impact of
channel-induced signal degradation and noise, which can decohere transmitted qubits, constrain the repetition rate of networking protocols, and limit the network's compatibility with various technologies~\cite{Cortes2022}. For example, timing drifts can degrade time-bin qubit signals and limit the synchronization of distributed systems. Additionally, optical phase fluctuations result in optical frequency broadening that impedes the integration of narrow-band technologies such as atomic-type quantum systems. Finally, drifting polarization can degrade polarization qubits and limit the performance of polarization-sensitive components.

Here we report on measurements of these characteristics for a fiber network connecting MIT Lincoln Laboratory (MIT-LL), MIT, and Harvard University, forming a Boston-Area Quantum Network (BARQNET). Based on this characterization, we design and demonstrate a compensation system that enables the transmission of photonic time-bin pulses from MIT-LL across the fiber to Harvard with $97.7\%$ fidelity. We discuss the future use of this compensation system and the BARQNET's potential for supporting emerging quantum networking protocols.

\begin{figure*}
    \centering
    \includegraphics{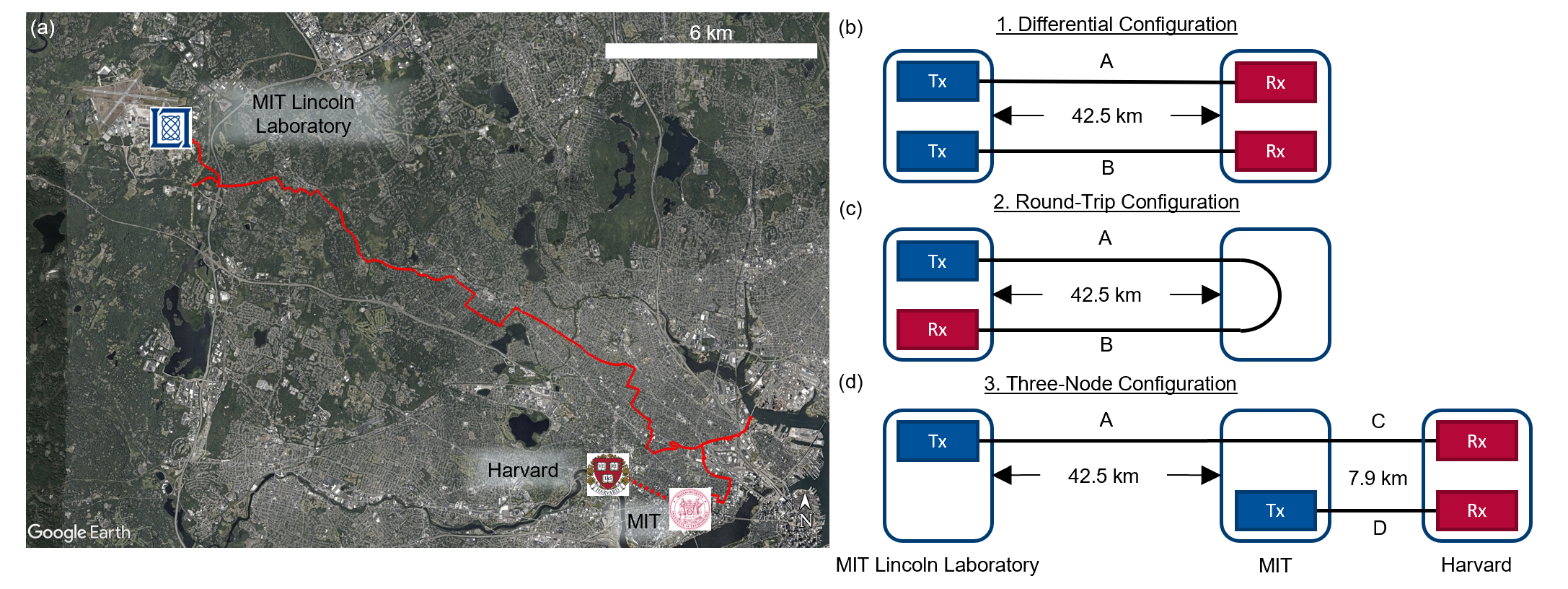}
    \caption{(a) Schematic of the BARQNET fibers connecting MIT Lincoln Laboratory in Lexington, MIT in eastern Cambridge, and Harvard University in central Cambridge. The solid red line shows a $\sim$40~km segment where the exact route is known, and the dashed line shows the final portion where the exact route could not be obtained. (b--d) Three different connectivity topologies explored in this work. The two copropagating fibers connecting MIT-LL and MIT are labeled A and B, and the two copropagating fibers connecting MIT and Harvard are labeled C and D. 
    }
    \label{fig:network}
\end{figure*}

Figure~\ref{fig:network}(a) depicts the BARQNET layout. Two 42.5~km spans of single-mode telecom fibers connect MIT-LL in Lexington, MA to MIT's main campus in Cambridge, MA. From there, an additional two 7.9~km spans of single-mode telecom fibers connect MIT to Harvard University, also in Cambridge, MA. All fibers are commercial plants, comprising various combinations of commercial SMF-28 fiber and non-zero dispersion shifted fiber that have been spliced together. Some of the fiber is buried, while other spans are above-ground on telephone poles. As a result, the link performance is below the ideal achievable using a laboratory spool or a custom-laid fiber connection, but is instead representative of the performance expected over existing and available deployed fiber links that will make up many emerging quantum network testbeds.

The three-node layout permits access to a number of configurations for both characterizing the fiber and performing quantum networking demonstrations. Fig.~\ref{fig:network}(b--d) depict three configurations explored in this work. The fibers can be used independently as in the Differential Configuration (Fig.~\ref{fig:network}(b)); this can be used to send quantum information down one fiber and classical information down another to avoid crosstalk~\cite{Chapuran2009,Burenkov2023}, or both can be used for quantum information to simply increase the channel bandwidth. Alternatively, connecting the fibers in the Round-Trip Configuration (Fig.~\ref{fig:network}(c)) provides a single 85~km span for testing protocols over longer distances. This configuration is also comparable to what would be used for two-way quantum information processing protocols, where a client sends data to a server which then returns a reply~\cite{Barz2012, Quan2022}. Finally, the additional fibers to Harvard can be utilized to realize a Three-Node Configuration  (Fig.~\ref{fig:network}(d)), whereby both MIT and MIT-LL can transmit information to Harvard, which in this topology serves as a central node. This configuration can be used for a range of demonstrations, such as single-repeater protocols~\cite{Braunstein_2012,Lo2012, Yin2016, Lucamarini2018,Bhaskar2020, Wang2022-4} or multipartite entanglement generation~\cite{Joshi2020, Pompili2021}.

\section{Deployed Fiber Characterization}

\begin{table}[b]
    \centering
    \begin{tabular}{|c|c|C{1.9 cm}|C{1.9 cm}|}\hline
        Fiber Span & Length (km) & Loss at 1550~nm~(dB) & Loss at 1350~nm~(dB)\\ \hline
        A & 42.5 & 11.9 & 16.6  \\ \hline
        B & 42.5 & 17.0 & 21.9  \\ \hline
        C & 7.9 & 10.4 & 11.2  \\ \hline
        D & 7.9 & 6.2 & 7.4  \\ \hline
    \end{tabular}
    \caption{Nominal length and fiber loss for each BARQNET span at wavelengths relevant for quantum networking with diamond silicon-vacancy centers. We note that these losses are primarily due to splices in this commercial fiber link rather than the expected inherent exponential propagation loss of $\sim0.35$~dB/km at 1350~nm and $\leq0.22$~dB/km at 1550~nm \cite{Corning_smf28}.}
    \label{tab:loss}
\end{table}

Photons exhibit a number of different degrees of freedom that can be used as an encoding basis for transmitting quantum information. In particular, photon number~\cite{Humphreys2018}, polarization~\cite{vanLeent2022}, and time-bin~\cite{Hensen2015} encodings have seen wide-spread exploration for applications such as remote entanglement generation and quantum key distribution. Transit over deployed optical fiber imparts noise that will affect each of these modalities in different ways, making it critical to understand the magnitudes and frequencies of these noise sources.

For example, modeling the performance of networked quantum systems with narrow atomic-resonance-based spectral acceptance windows can benefit from characterizing fiber-induced optical linewidth drift and broadening down to kHz-class resolution~\cite{Zhong_2018}. Additionally, predicting the performance of polarization-sensitive components calls for characterizing fiber-induced polarization drift to single degrees, for example to keep error rates below 1\% for polarization-encoded qubits. Lastly, projecting how desired GHz-rate communication protocols would perform requires characterizing fiber-induced timing jitter to below one nanosecond.

This sort of characterization requires a stable and accurate common reference between nodes. This is straightforward for polarization by using the direction of the force of gravity to define a common vertical axis, but is more challenging for time and frequency. Easily accessible microwave synchronization signals such as GPS do not offer the sub-nanosecond timing precision needed~\cite{Vyskocil2009}, and optical systems like the White Rabbit Protocol are typically limited to 10~km~\cite{WhiteRabbit}. High precision optical frequency distribution systems are, at present, not feasible for use as a portable channel characterization system~\cite{Lopez2012,Predehl2012,Newbury2016}. Instead, we make use of the BARQNET's reconfigurability to characterize the fiber-induced degradation between nodes using only local references.

Operating in the Round-Trip Configuration --- where a characterization signal is sent along one fiber, applying a noise operator $\hat{A}$, and returned along another, applying a noise operator $\hat{B}$ --- enables the returned signal to be directly compared to the local reference which allows the combined effect of both fiber spans $\hat{A}\cdot\hat{B}$ to be measured. Alternatively, operating in the Differential Configuration --- where a characterization signal is sent along one fiber and a reference signal is sent along the other fiber --- enables measurement of the correlation of the two fibers' transforms $\hat{A}$ and $\hat{B}$. These sum and difference characterizations allow us to determine the expected effect of a single fiber span in our testbed. Table~\ref{tab:loss} provides the loss experienced over each span used for all configurations at 1550~nm and 1350~nm, the two wavelengths examined in this work due to their relevance for networking with diamond silicon-vacancy centers.

\subsection{Optical phase and frequency noise characterization}
Quantum communication schemes that utilize a photon-number-based encoding require links with known phase differences, necessitating active stabilization or tracking of the phase delay imparted by the optical fiber on the signal~\cite{Cabrillo_1999,Minar2008, Sangouard2011}. Drifts in phase can also impart Doppler shifts on transported light, inducing a frequency shift that can degrade the coherence of sufficiently narrow bandwidth signals. We characterize this phase noise in both the differential and round-trip configurations shown in Fig.~\ref{fig:network}(b--c). For this measurement, the transmitter is a single long-coherence-length laser, split by a 50:50 beamsplitter to permit comparison of the different transit paths. For the Differential Configuration, these two paths are the two fibers; for the Round-Trip Configuration, one path is the full 85-km fiber loop, while the other is a short ($<1$~m) local span of fiber. For this measurement, the receiver employs an optical pi-hybrid device that performs an interferometric measurement of the optical phase difference $\Delta\phi(t)$ between two input paths.

\begin{figure}
    \centering
    \includegraphics{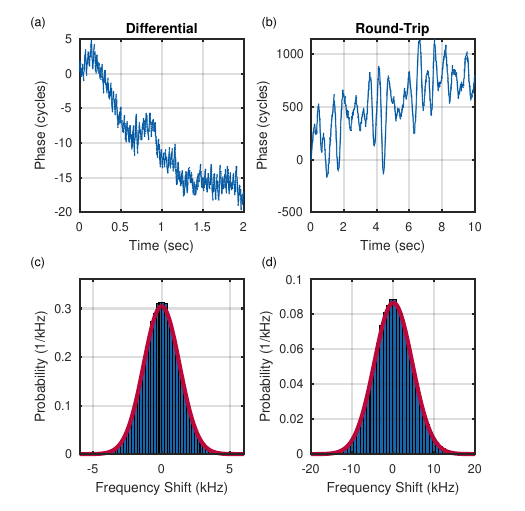}
    \caption{Measurement of phase drift over the deployed link in both (a) Differential and (b) Round-Trip Configurations, downsampled to 50~kHz to remove noise from our measurement equipment. Differentiating these data provides a measurement of frequency shift over time. Due to the random nature of these fluctuations, this shift is effectively a frequency broadening, with a profile given by the histograms shown in (c) and (d), which are fit to Gaussian profiles (red) with variances $V_D=1.72$~kHz$^2$ and $V_R=21.2$~kHz$^2$ for the Differential and Round-Trip data respectively.
    }
    \label{fig:phase}
\end{figure}

Figure~\ref{fig:phase} shows the raw phase drifts as measured in the Differential (Fig.~\ref{fig:phase}(a)) and Round-Trip (Fig.~\ref{fig:phase}(b)) Configurations, downsampled to 50~kHz to remove noise from our measurement equipment (see Supplemental Material~\cite{supplement} for dates and times of all fiber characterization data). In both cases, the phase difference fluctuates across many full $2\pi$ cycles over the course of a second. These findings are orders of magnitude larger compared to results on buried fibers~\cite{Minar2008,Amies-King2023}, demonstrating the critical importance of the fiber environment to the diffusion process. Due to the BARQNET's fast phase fluctuations, any phase-sensitive quantum communication protocol would require high bandwidth stabilization.
For example, a control circuit with a 3-dB bandwidth of 650~kHz and 39$^\circ$ phase margin was used to stabilize the BARQNET to a residual phase noise of \num{3e-2} radians~\cite{Grein2017}. 

Differentiating the phase drift signal provides a measurement of the Doppler-induced frequency shift over time: $\Delta f = d \Delta \phi(t)/dt.$ Due to the random nature of these fluctuations, this shift is effectively a frequency broadening, with a profile given by the histograms shown in Fig.~\ref{fig:phase}(c) and Fig.~\ref{fig:phase}(d) for the Differential and Round-Trip Configurations respectively. We model the probability density function of an optical frequency shift $\rho(\Delta f)$ on a given fiber as resulting from a Brownian process:
\begin{equation}
\rho(\Delta f) = \left(\frac{1}{\sqrt{2\pi V}}\right) e^{-\Delta f^2 /2 V},
\end{equation}
which has variance $V$. Our measurements then correspond to the sum ($\rho_{R}$ for the Round-Trip Configuration) and difference ($\rho_{D}$ for the Differential Configuration) of the distributions $\rho_{A}$ and $\rho_{B}$, which have a covariance $C$.
From the data in Fig.~\ref{fig:phase}(c--d), we can calculate the amount of common-mode noise due to the two strands' co-propagation. Using the measured variance of the differential optical frequency noise, $V_D = 1.72$~kHz$^2$, and the measured variance of the round-trip optical frequency noise, $V_R = 21.2$~kHz$^2$, along with the assumption that the optical 
phase
noise contributions from each single 42.5~km fiber span have equal magnitude $V_{\hat{A}} = V_{\hat{B}} \equiv V$, we can calculate the variance of the optical frequency noise for each fiber span $V = \left( V_R + V_D \right)/4 = 5.74$~kHz$^2$, as well as the covariance of the optical frequency noise for the two fiber spans  $C = \left( V_R - V_D \right)/4  = 4.88$~kHz$^2$.  
The relatively strong covariance suggests that quantum networking applications requiring phase-stable links can benefit from distributing phase references between nodes, even if the reference is distributed in a separate but co-propagating optical fiber. Importantly, in all cases, the effective frequency broadening is narrower than the bandwidth of typical quantum networking systems, which range from $10$~MHz-class~\cite{Hensen2015,vanLeent2022} to $10$~GHz-class~\cite{Woodward2021}.

The good agreement between our data and the Brownian model suggests phase drifts are well-described by a diffusive process. In this case, the variance $V$ is linearly proportional to the length of the fiber $L$. Our result can thus be scaled to other fiber lengths in similar environments by using the variance per unit length $v=133~$Hz$^2$/m.

\subsection{Polarization drift characterization}

Polarization is an attractive degree of freedom for encoding photonic qubits due to its ease of generation, manipulation, and measurement. Furthermore, as a spatial degree of freedom, polarization measurements can in principle be absolute without requiring distribution of a reference across network nodes. However, strain-induced birefringence in optical fibers can lead to polarization transformations which drift over time as the fiber environment changes~\cite{Agrawal2010}. We characterize the drift in polarization transforms $\hat{A}_{p}$ and $\hat{B}_{p}$ over the BARQNET, specifically measuring the drift in the polarization sent in the Differential Configuration (in this case only sent one-way along fiber A from MIT to MIT-LL), as well as in the Round-Trip Configuration  (sent two-way from \mbox{MIT-LL} to MIT and back). Our transmitter is once again a single long-coherence-length laser, and our receiver is a polarization analyzer which records the polarization vector $\hat{p}$ in terms of the Stokes parameters $\{S_1,S_2,S_3\}$ at a rate of 1~Hz. 

\begin{figure}
    \centering
    \includegraphics[trim={0 1cm 0 0.5cm}]{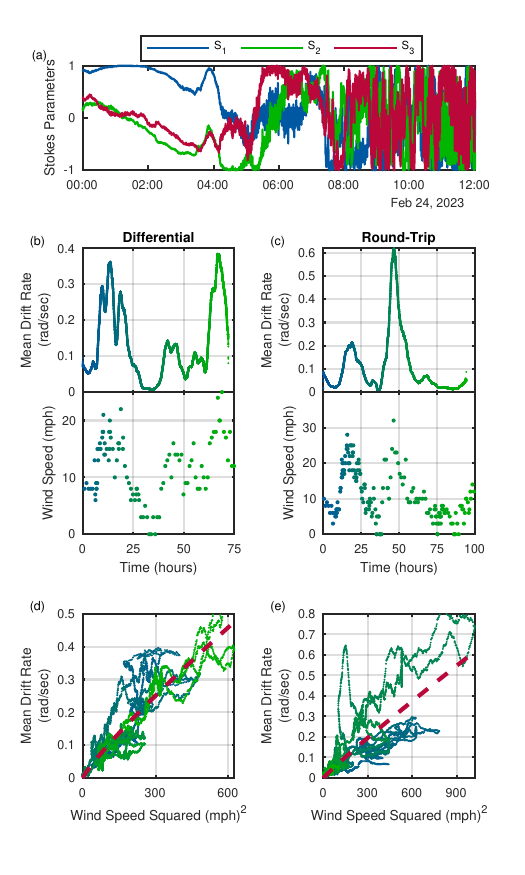}
    \caption{Measurement of polarization drifts. (a) Example trace of polarization drift measured in the differential configuration, showing the Stokes parameters over a typical 12 hour period. The 10 minute rolling average of this drift is calculated for (b) the Differential Configuration and (c) the Round-Trip Configuration, each compared with the concurrent average wind speed recorded at MIT-LL. The correlations between drift rate and the square of the wind speed are plotted against one another in (d--e), and fit with an exponential dependence (red).  Point colors in (d) and (e) correspond to the like-colored time-points in (b) and (c), respectively.}
    \label{fig:pol}
\end{figure}

Figure~\ref{fig:pol}(a) shows a characteristic trace of the polarization over twelve hours, plotting the Stokes parameters as measured by our polarization analyzer. We observe high variability in the drift rate, with long periods of stability as well as periods of rapid fluctuation. To better understand the source of this variance, we measured the polarization over the course of multiple days in both configurations. We then calculate the drift in polarization angle between successive measurements (spaced 1~second apart), and determine a mean polarization angle drift rate $\langle\dot{\Theta}\rangle$ by taking a rolling average over 10~minutes. 
Figure~\ref{fig:pol}(b--c) show the measured $\langle\dot{\Theta}\rangle$ as a function of time for the Differential and Round-Trip Configurations respectively, alongside the average wind speed $W$ recorded outside of MIT-LL~\cite{wunderground}, showing a clear correlation between the two datasets. 
We plot these correlations between $\langle\dot{\Theta}\rangle$ and the square of the wind speed $W^2$ in Figure~\ref{fig:pol}(d--e), where we have linearly interpolated the wind speed data to match the sampling rate of the polarization.

These data are fit (red dashed line) using a power law ${\langle\dot{\Theta}\rangle = \kappa \times W^{n}}$.  For the one-way data taken using the Differential Configuration, the fit gives ${\kappa_{D} = 1.74(5)~\textrm{mrad}/\textrm{sec}\cdot(\textrm{mph})^{n_{D}}}$ and ${n_{D} = 1.74(1)}$ with an adjusted $R^2=0.441$. For the two-way data taken using the Round-Trip configuration, the fit gives ${\kappa_{R} = 0.94(3)~\textrm{mrad}/\textrm{sec}\cdot(\textrm{mph})^{n_{R}}}$ and ${n_{R} = 1.87(1)}$ with an adjusted $R^2=0.420$.  
Under a Brownian model, we expect $\langle\dot{\Theta}\rangle$ to depend quadratically on the wind speed, which is consistent with these results (see Supplemental Material~\cite{supplement} as well as Ref.~\cite{Joffre_1988} therein).
To understand the slight reduction of $\kappa_{R}$ relative to $ \kappa_{D}$, we can consider that while the one-way measurement probes the transformation $\hat{A}_{p}\hat{p}$, the two-way measurement probes $\hat{B}_{p}^{-1}\hat{A}_{p}\hat{p}$. The modest reduction here thus suggests some common-mode polarization drift between the two fibers, such that $\hat{A}_{p}$ is partially counteracted by $\hat{B}_{p}^{-1}$.

These results indicate that the specific environment of a given deployed fiber network will have a strong impact on the polarization drift rate. Indeed, while our figures are comparable to those reported for links of similar length and with similar amounts of unburied fiber~\cite{Dixon2015}, other demonstrations using fully-buried fiber have observed significantly reduced polarization drift~\cite{Amies-King2023}.

\subsection{Optical path length drift characterization}

\begin{figure}
    \centering
    \includegraphics{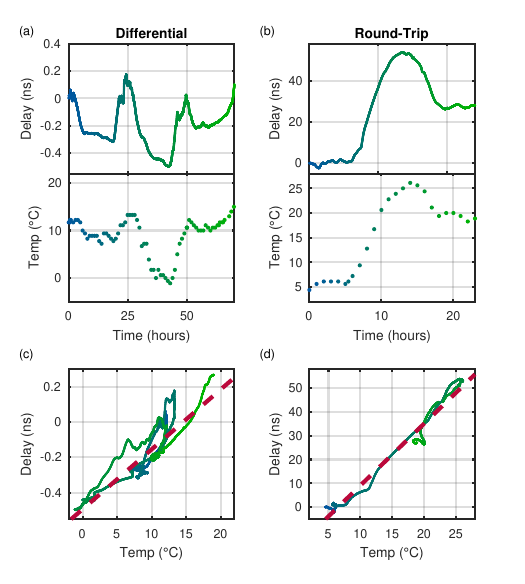}
    \caption{Optical path length drifts. Time-of-flight measurements in the (a)  Differential Configuration measured relative to the nominal delay $t_D=108.4$~ns, and in the (b) Round-Trip Configuration measured relative to the nominal delay $t_R=415.045$~\textmu s. For each we also plot the temperature measured concurrently outside of MIT-LL. The resultant correlation yields a linear fit (red dashed line) with a dependence of (c) 33.9(2)~ps/$^\circ$C for the Differential Configuration and (d) 2.59(1)~ns/$^\circ$C for the Round-Trip Configuration. Point colors in (c) and (d) correspond to the like-colored time-points in (a) and (b), respectively.}
    \label{fig:delay}
\end{figure}

Finally, we consider the drift in optical path length over the BARQNET, as measured by photon time-of-flight. Time synchronization between nodes is a core capability for any quantum network, in order to distinguish between temporally-multiplexed classical and quantum communication time bins, and in general to coordinate the different stages of protocols. This capability is particularly important when encoding photonic qubits in the time-bin basis~\cite{Marcikic2002}, which can require sub-nanosecond timing synchronization precision~\cite{Lee2019}. 
The time-bin basis has seen popularity due to its resilience to noise; because all time-bins will travel along a common path, any phase noise in-transit will be global and thus not decohere the qubit~\cite{Barrett_2005,Hensen2015}. Furthermore, using time-bins with a single polarization and applying polarization filtering can convert such drifts from bit errors into efficiency losses. Since most networking protocols have stronger constraints on bit errors (typically $<10\%$) than on losses, this can improve overall protocol performance using only modest polarization drift correction. 

The time-of-flight $\tau$ across an optical fiber depends on the length of fiber $L$ and the optical index of refraction of the fiber $n$, which are both dependent on the temperature $T$. As a result, changes in the path length are given by:

\begin{equation}
\Delta \tau = \left( \alpha_{L} + \alpha_{n}  \right) \tau_{0} \Delta T,
\end{equation}
where $\alpha_{L}$ is the fiber's linear thermal expansion coefficient, $\alpha_{n}$ is the temperature dependence of the fiber's index of refraction, and $\tau_0$ is the nominal time-of-flight across the fiber.
We characterize the drift in $\tau$ over the BARQNET by measuring the time-of-flight across the fiber. In the Differential Configuration, we compare the difference in transit time across links $A$ and $B$, which has a nominal value of $\tau_D=108.4$~ns. In the Round-Trip Configuration, we measure the total time-of-flight referenced to transit across a short ($<1$~m) local span of fiber, giving a nominal round-trip time of $\tau_R=415.045$~\textmu s. In both cases, our transmitter is a pulsed laser source generated by electro-optic intensity modulation of a continuous wave laser, split as in the phase drift experiment to enable comparison between the two transit paths. Our receiver is a fast photodiode connected to a time tagger which records the difference in arrival time between pulsed signals traveling each path.

We record the delay in transit between the two fibers measured in the Fig.~\ref{fig:delay}(a) Differential Configuration relative to $\tau_D$, and the Fig.~\ref{fig:delay}(b) Round-Trip Configuration relative to $\tau_R$. Below this, we plot the temperature measured outside of MIT-LL.  Figure~\ref{fig:delay}(c--d) shows the resultant correlation, giving a linear fit of 33.9(2)~ps/$^\circ$C (adjusted $R^2=0.790$) and  2.59(1)~ns/$^\circ$C (adjusted $R^2=0.965$) for the Differential and Round-Trip Configurations respectively.

The temperature dependence coefficients have expected values of ${\alpha_L\approx}$~\num{0.5e-6}$/^\circ$C~\cite{
%Gilmore2014, 
Cavillon2017} and \mbox{${\alpha_n\approx}$~\num{8e-6}$/^\circ$C~\cite{Leviton_2006}}. Using these numbers and assuming that our two fibers are nearly identical in terms of nominal values $\tau_0$ as well as temperature experienced at any given time predicts a dependence of 3.6~ns/$^\circ$C over the 85~km Round-Trip Configuration. The closeness of this value with our measured result indicates that the temperature change measured at one end of the network is very close to that experienced by the fiber along its entire span, despite many sections being underground. This suggests that for networks with multiple nodes within tens of kilometers, the temperature at a single node can be used to predict the behavior of optical path length for fibers throughout the network. Furthermore, the suppression of path length drift by nearly two orders of magnitude for the Differential Configuration reveals that the assumption of identicality between the two fibers is reasonable to single percent levels of precision. 

\section{Time-Bin Qubit Distribution Protocol}

\begin{figure*}
    \centering
    \includegraphics{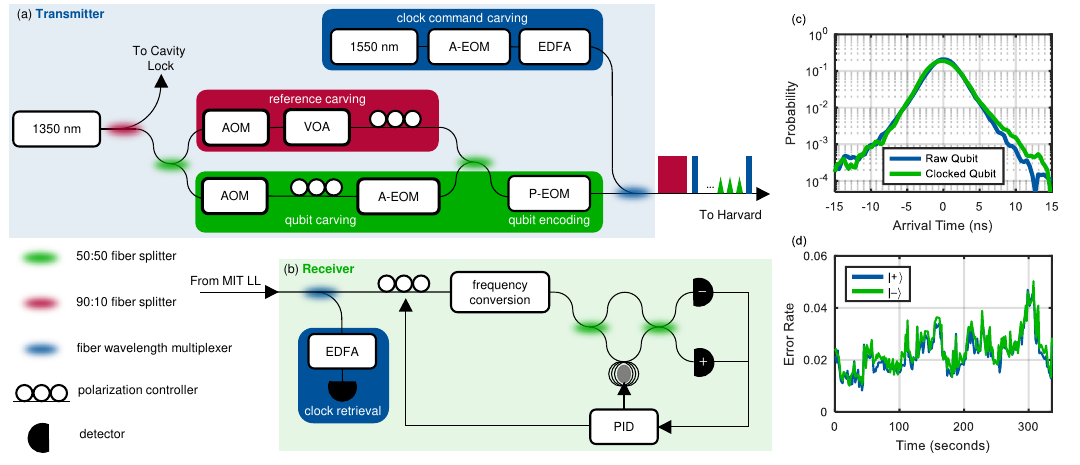}
    \caption{Optical setup for memory-compatible time-bin qubit (a) transmitter and (b) receiver. A-EOM = amplitude electro-optic modulator, P-EOM = phase electro-optic modulator, AOM = acousto-optic modulator, EDFA = erbium-doped fiber amplifier, VOA = variable optical attenuator, PID = proportional-integral-derivative feedback controller.} (c) Clock distribution from Alice to Bob. The shape of our 1350~nm qubit pulses retrieved from autocorrelation measurements (blue) is nearly identical to the shape retrieved when triggering off our timing signal (green), with only a 520~ps broadening in time-of-arrival variance. (d) Error rate of photonic qubit preparation and measurement in the $X$-basis over a 336~second transmission period. Since our photon-to-spin mapping scheme requires measurement of photonic qubits in the $X$-basis, we characterize our encoding in this basis by sending $\ket{+}$~(blue) and $\ket{-}$~(green). This gives a mean state preparation and measurement error rate of 2.3\%. We attribute the increase in error rate over time to drifts in the TDI polarization, which decrease interference contrast.
    \label{fig:setup}
\end{figure*}

Cognizant of the above noise sources experienced by light on our deployed fiber link, we design a communication scheme for transmitting quantum information that is resilient to the fastest noise sources, permits compensation for slower noise sources, and is compatible with the operation of a quantum memory on the receiving end. Specifically, we target a protocol for quantum state transfer in the three-node configuration from Fig.~\ref{fig:network}(b), where time-bin photonic qubits are sent from MIT-LL (Alice) and received and stored by a nanocavity-coupled silicon-vacancy (SiV) center in diamond~\cite{Nguyen2019, Nguyen2019b} at Harvard (Bob). With the addition of a third node such as a second transmitter at MIT, this protocol has utility for example in memory-enhanced quantum communication~\cite{Bhaskar2020}. For compatibility with the SiV quantum memory, we incorporate a frequency conversion setup between telecom and visible wavelengths, described in Ref.~\cite{Bersin2023}.

The transmitter waveform for our scheme is depicted in Figure~\ref{fig:setup}(a), and comprises a classical clock signal at 1550~nm for timing synchronization, a quantum signal at 1350~nm containing our photonic qubits, and a classical reference signal at 1350~nm for fiber noise compensation. Our photonic qubits are generated by attenuated coherent state pulses. Light from a 1350~nm laser is split out, passed through an acousto-optic modulator (AOM) for power and frequency control, then passed through an amplitude electro-optic modulator (A-EOM) which carves pulses to form time bin qubits. Each pulse is a Lorentzian with full width at half maximum $\Gamma=45$~ns, and the time bins are spaced 144.5~ns apart, set to be compatible with the low-bandwidth constraints of the SiV quantum memory~\cite{Bhaskar2020}. After pulse carving, a phase EOM (P-EOM) encodes phases on each time bin. At the receiver (Fig.~\ref{fig:setup}(b)), these qubits are upconverted to 737~nm using sum-frequency generation facilitated by a 1623~nm pump laser~\cite{Bersin2023} then measured on a time-delay interferometer (TDI) matched to the spacing of our time-bins.

For timing synchronization, 
optical  systems like the White Rabbit (WR) protocol~\cite{WhiteRabbit} offer a flexible means of synchronizing network timing with 100~ps-class precision. However, the continuous-transmission nature of the WR protocol demands high isolation to prevent cross-talk between timing data and the co-propagating, single-photon-level qubits. Furthermore, our network's high loss is prohibitive for standard SFP transceivers, which can typically tolerate no more than 20~dB of loss.
We thus develop a custom optical sequence that uses low-duty cycle optical pulses to both minimize quantum-classical cross-talk by avoiding temporal overlap as well as lower the average power necessary for achieving synchronization. We transmit 1550~nm optical optical pulses (duration 100~ns) in a sequence that
comprises a trigger pulse, followed by a series of pulses (repetition rate 5~MHz) that communicate what portion of the experimental sequence should be performed, such as data transmission or fiber noise compensation. These ``clock command'' words are distinct from one another by at least two symbols, allowing us to avoid sequence errors from packet loss. Such packet loss errors are in practice rare due to our robust amplification scheme. Timing pulses are amplified to 14~dBm with an erbium-doped fiber amplifier (EDFA) 
%(MPB) 
before transmission, and are amplified to 6~dBm with a second EDFA on receipt. The detected signal is then further amplified with a pulse amplifier 
%(iXblue) 
to 5~V, providing a reliable, low-jitter trigger signal. Comparing the natural pulse shape of our qubits with the pulse shape retrieved by triggering off our optical clock (Fig.~\ref{fig:setup}(c)), we measure only 520~ps broadening in time-of-arrival variance, well below the duration of our qubit time-bins.
This enables us to achieve nanosecond-level timing precision across our network despite the $>200$~\textmu s transit time between each site, as well as perform on-the-fly coordination between transmitter and receiver sequencers~\cite{supplement}.

Finally, we also periodically distribute longer reference pulses at 1350~nm, which serve two purposes. First, since our qubits are encoded in the time-bin degree of freedom, fiber-induced polarization errors do not reduce the fidelity of our protocol. However, due to the polarization-dependent nature of other receiver components such as frequency conversion, single-mode nanophotonic devices, etc., such drifts manifest as loss. A common solution is to perform periodic, automated polarization correction based off a strong pilot tone of known polarization~\cite{Rosenfeld2008,Yu2020,Du2021}. We periodically send a 10~second reference pulse of continuous 1350~nm light, during which time an automated sequence rotates waveplates to compensate for polarization drifts, 
maintaining the desired polarization to within $<20^\circ$.
Second, the TDI used to measure our photonic qubits has a small FSR around 7~MHz. As a result, drifts in our 1350~nm or 1623~nm laser frequencies $\geq$100~kHz or drifts in the TDI path length imbalance $\geq$10~nm will cause percent-level phase errors that degrade our protocol fidelity. We passively dampen these drifts by locking these two lasers to ultra low expansion Fabry-P\'{e}rot cavities
to ensure narrow linewidths ($<100$~kHz)
and by placing our TDI in a thermally and vibrationally isolated case. We also actively adjust the length of the TDI to compensate for any slow drifts. We periodically send a 1~second reference pulse of continuous 1350~nm light, during which time a locking sequence applies a voltage to stretch and compress a piezo spool in the long arm of the TDI. This locking signal is gated by an acousto-optic modulator (AOM), allowing us to shift the frequency of the lock signal relative to our qubit frequency. In this way, our feedback sequence locks the TDI to the quadrature point of the reference light such that our incoming qubits will interfere at the fringe maxima. The piezo voltage is then held during the portion of the sequence when qubits are sent. A variable optical attenuator (VOA) is used to increase extinction on the reference signal during qubit transmission periods.

To benchmark the efficacy of our overall communication system, we perform a send-and-measure experiment which is identical to the protocol performed for mapping photonic qubits to an SiV~\cite{Bhaskar2020,Bersin2023}. To test the full system of timing synchronization, polarization correction, and frequency distribution, Alice sends a string of qubits encoded in the $X$-basis $\{\ket{+},\ket{-}\}$, along with all of the aforementioned timing and reference signals. Here $\ket{+}$ corresponds to a photon split equally into early and late time bins, while $\ket{-}$ is the same state but with a $\pi$ phase shift on the late bin. Figure~\ref{fig:setup}(d) shows the result of measuring these qubits after they have traversed the full system --- traveling across the 50~km deployed fiber link from Alice to Bob, undergoing polarization correction, being upconverted, and interfering on the TDI. To isolate the impact of our cross-fiber synchronization from other noise sources like detector dark counts and noise from the frequency converter, we send an increased number of photons per pulse compared to a true secure QKD protocol, measuring mean photon number $\langle \hat{n}\rangle= 0.0202(5)$ at the point of detection. As shown in Fig.~\ref{fig:setup}(d), we retrieve a mean bit-error-rate of 2.3(6)\% across the $\ket{+}$ and $\ket{-}$ states, where the error rate is defined as the probability to measure the opposite of the transmitted bit. We attribute the remaining error to imperfect interference visibility of our TDI due to internal polarization drifts, which was observed to degrade over the timescale of this measurement, likely due to air currents and human movement within the room. Importantly, this can be corrected by having a local polarization reference and correction scheme without modifying the cross-fiber transmission protocol.

\section{Discussion}
Our results demonstrate that the BARQNET's fundamental noise characteristics are suitable for a range of future experiments.
The low fiber-induced optical frequency noise indicates compatibility with narrow-bandwidth atomic-resonance-type quantum systems down to kHz-class with all-local frequency stabilization and referencing.
Its nanosecond-class time-of-flight variations occur slowly and predictably based on weather conditions,
indicating that even with Hz-class cross-fiber synchronization, the BARQNET can support up to MHz-class quantum network clock rates. Its polarization drift rates, even during high-wind periods, average below 1~rad/sec, indicating that high fidelity polarization stabilization can be achieved with $\sim$10-kHz-class
stabilization system.
These low-bandwidth requirements for the classical support infrastructure leave the majority of the channel bandwidth available, allowing high duty cycle transmission of quantum signals. 

The experiments reported here constitute a comprehensive characterization of the relevant figures of merit for quantum networking. Importantly, the application of these characterization techniques enables us to design a high-fidelity photonic distribution system that is resilient to noise across the BARQNET.
These techniques are of significance for the development of emerging quantum network testbeds~\cite{Valivarthi2020, Cui2021, Alshowkan2021, Du2021, Luo2022, Kapoor2023}, helping to both reveal what demonstrations are presently feasible and identify the most impactful near-term improvements to both the quantum components and classical infrastructure.
For example, our results indicated that the fiber-induced phase and frequency noise, polarization drift, and path length drift characteristics of our link are compatible with performing many essential networking protocols over our testbed. Indeed, we used these findings to guide the design of a system that successfully integrates quantum memories~\cite{Bersin2023}. While more advanced demonstrations such as entangling two memories across the BARQNET are at present impeded by the high loss on fibers $A$ and $B$, these loss characteristics can be improved by replacing the anomalously high loss fiber splices and cross-connects in our links, in principle lowering the loss from MIT-LL to Harvard down to the propagation loss limit near 10~dB. With these considerations in mind, we are currently working to interface multiplexed quantum memories at each node~\cite{Starling2023} as well as incorporating additional nodes into the BARQNET for system-level exploration of quantum networking protocols.

\begin{acknowledgments}
This material is based upon work supported by the National Reconnaissance Office and the Under Secretary of Defense for Research and Engineering under Air Force Contract No. FA8702-15-D-0001. Any opinions, findings, conclusions or recommendations expressed in this material are those of the authors and do not necessarily reflect the views of the National Reconnaissance Office or the Under Secretary of Defense for Research and Engineering. \textcopyright~2023 Massachusetts Institute of Technology. Delivered to the U.S. Government with Unlimited Rights, as defined in DFARS Part 252.227-7013 or 7014 (Feb 2014). Notwithstanding any copyright notice, U.S. Government rights in this work are defined by DFARS 252.227-7013 or DFARS 252.227-7014 as detailed above. Use of this work other than as specifically authorized by the U.S. Government may violate any copyrights that exist in this work.

We thank Franco N. C. Wong and Ian Christen for helpful discussions. This work was supported by the National Science Foundation (NSF, Grant No. PHY-2012023), NSF EFRI ACQUIRE (Grant No. 5710004174), Center for Ultracold Atoms (Grant No. PHY-1734011), Department of Energy (DoE, Grant No. DESC0020115), AFOSR MURI (Grants No. FA9550171002 and No. FA95501610323), and the NSF ERC Center for Quantum Networks (Grant No. EEC-1941583). E.B. and M.S. acknowledge funding from a NASA Space Technology Research Fellowship. Y.Q.H. acknowledges support from the Agency for Science, Technology and Research (A*STAR) National Science Scholarship. D.A., E.N.K., and B.M. acknowledge support by the NSF Graduate Research Fellowship under Grant No. DGE1745303. R. R. acknowledges support by the Alexander von Humboldt Foundation, the Cluster of Excellence `Advanced Imaging of Matter’ of the Deutsche Forschungsgemeinschaft (DFG) - EXC 2056 - project ID 390715994, BMBF Project 16KIS1592K.
\end{acknowledgments}

\bibliography{references}

\end{document}

%%%%%%% Begin Supplemental materials %%%%%%%

\clearpage
\widetext
\begin{center}
\textbf{\large Supplementary Information for ``\titlename''}
\end{center}
%%%%%%% Prefix a "S" to all equations, figures, tables and reset the counter %%%%%%%
\setcounter{equation}{0}
\setcounter{figure}{0}
\setcounter{table}{0}
\setcounter{section}{0}
\setcounter{page}{1}
\makeatletter
\renewcommand{\theequation}{S\arabic{equation}}
\renewcommand{\thefigure}{S\arabic{figure}}
\renewcommand{\thetable}{S\arabic{table}}
\renewcommand{\thesection}{S\arabic{section}}
\renewcommand{\bibnumfmt}[1]{[1]}
\renewcommand{\citenumfont}[1]{#1}
%%%%%%%%%% Prefix a "S" to all equations, figures, tables and reset the counter %%%%%%%%%%

\section{Data Collection Dates}

Table~\ref{tab:dates} provides the dates and times corresponding to $t=0$ for each fiber characterization dataset. All times are given in Eastern Time (ET). Exact times were unavailable for the phase measurements, but are known to have been during daytime work hours.

\begin{table}[h]
    \centering
    \begin{tabular}{ |c|c|c| }
    \hline
    Noise Process & Configuration & Datetime at $t=0$ \\ \hline
    \multirow{2}{*}{Phase} & Differential & 2019-08-28, Daytime \\
     & Round-Trip & 2017-04-06, Daytime \\ \hline
    \multirow{2}{*}{Polarization} & Differential & 2023-02-22, 13:36 \\
     & Round-Trip & 2023-03-03, 18:02 \\ \hline
    \multirow{2}{*}{Optical Path Length} & Differential & 2017-10-30, 12:44 \\
     & Round-Trip & 2016-03-09, 06:00 \\
    \hline
    \end{tabular}
    \caption{Starting dates and times for each characterization dataset, formatted \mbox{''Year-Month-Day, Hours:Minutes''}.}
    \label{tab:dates}
\end{table}

\section{Polarization Drift Model}
We model the fiber-induced polarization drift as resulting from a Brownian process. Here, the two-dimensional nature of polarization results in an {angular} drift rate $\dot{\Theta}$ that follows a Rayleigh probability distribution:

\begin{equation}
\rho_p(\dot{\Theta}) = \frac{\dot{\Theta}}{\sigma_p^2}e^{-\dot{\Theta}^2/2\sigma_p^2},
\end{equation}

which has non-zero mean $\langle\bar{\dot{\Theta}}\rangle=\sigma_p\sqrt{\pi/2}$. A possible mechanism for this drift is stress-induced birefringence, where the stress stems from force applied by the wind. In this case, the standard deviation of the drift $\sigma_p$ would stem from variations in the wind force, which have a standard deviation $\sigma_F$. As the force applied by a fluid to a surface is proportional to the fluid velocity squared, this implies proportionality to the standard deviation of the wind speed squared, $\sigma_F\propto \sigma_W^2$. Finally, $\sigma_W$ is reported to be linearly proportional to the wind speed $W$
%\cite{Joffre_1988}.
[47].
Thus, a model of Brownian motion caused by variations in the wind speed should exhibit a relationship $\langle\bar{\dot{\Theta}}\rangle=\kappa\times W^2$ with a scaling factor $\kappa$, which depends on physical parameters of the system such as the fiber stress responsivity, the area of the fiber bundle facing the wind direction, etc. 

\begin{figure}
    \centering
    \includegraphics{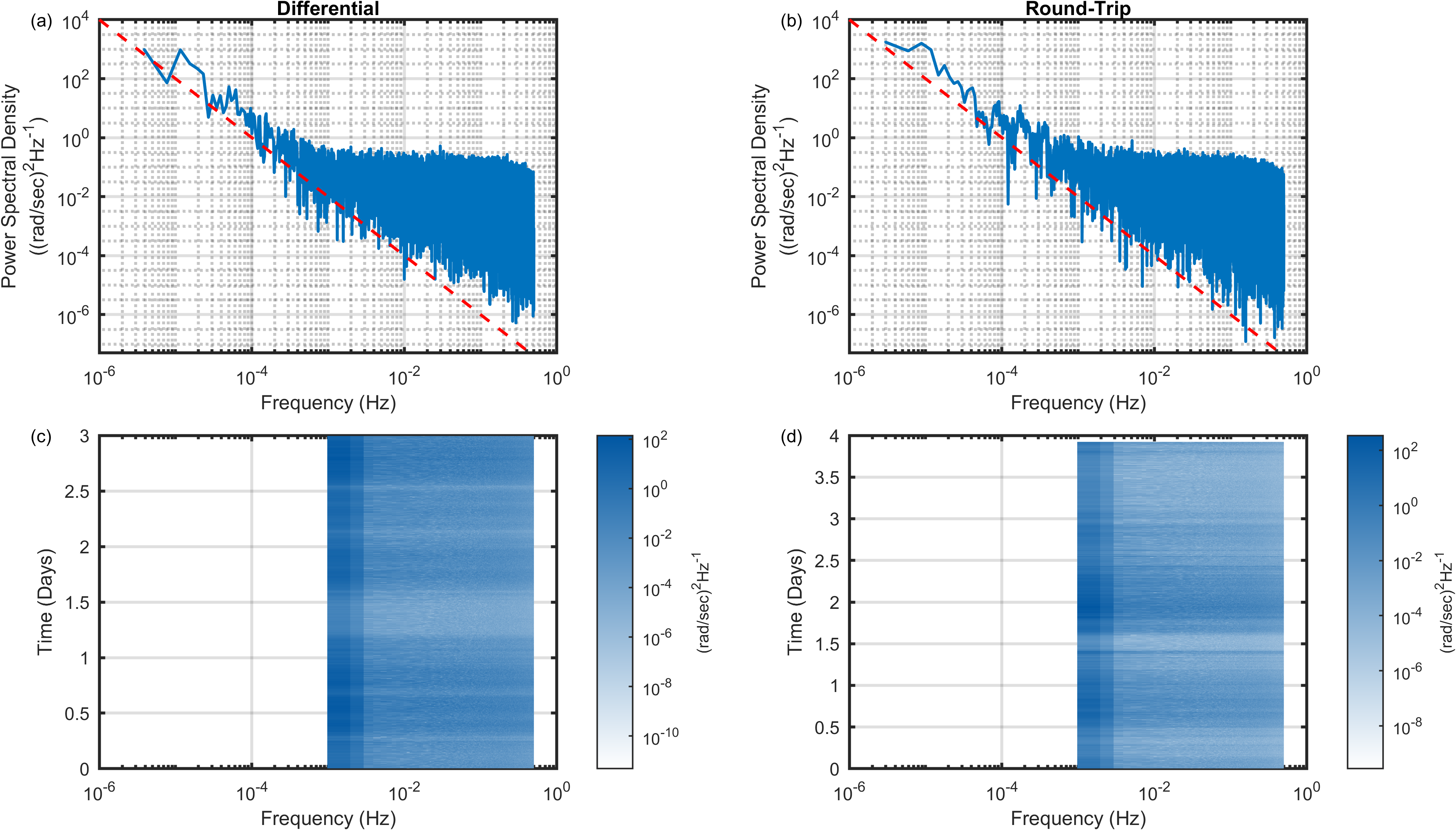}
    \caption{Power spectral density over the entire data collection period of $\dot{\Theta}$ for (a) the Differential and (b) the Round-Trip Configurations. {The red dashed lines are a reference 20~dB-per-decade expected for Brownian noise.} We also show the spectrograms to capture the time dynamics in (c) and (d) for the Differential and Round-Trip Configurations respectively.}
    \label{fig:pol_psd}
\end{figure}

Figure~\ref{fig:pol_psd} shows the power spectral density of the polarization drift rate $\dot{\Theta}$ for the (a) Differential Configuration and (b) Round-Trip Configuration, {along with a 20-dB-per-decade reference line (red dashed), the expected profile for noise caused by a Brownian process.} In both cases, this is taken over the entire time trace of the data, and thus does not capture the differing dynamics observed under different wind speed conditions. {Indeed, while the floor of the noise appears to follow the expected Brownian profile, there are large fluctuations in the spectrum at higher frequencies due to the variation in the spectrum across the time trace.} To explore the time-varying nature of this power spectral density, we also plot spectrograms for the (c) Differential and (d) Round-Trip Configurations, where in each case the signal has been binned into 10-minute windows. {Note that this binning limits the low-frequency resolution of the spectrograms to $\sim2$~mHz.}

\begin{figure}
    \centering
    \includegraphics{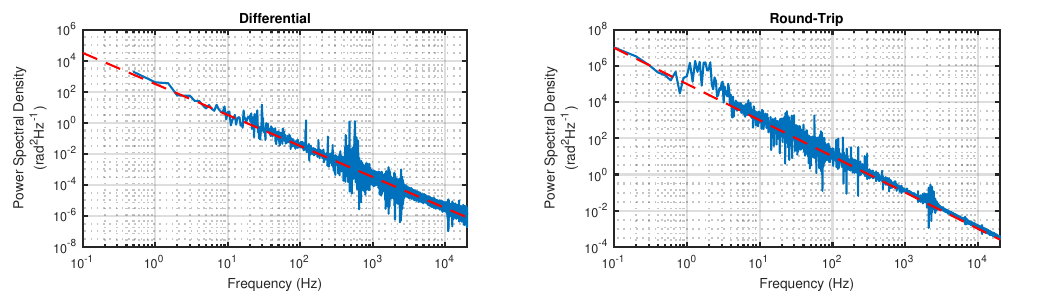}
\caption{{Power spectral density of phase drift for (a) the Differential and (b) the Round-Trip Configurations. The red dashed lines are a reference 20~dB-per-decade expected for Brownian noise.}}
    \label{fig:phase_psd}
\end{figure}
\begin{figure}
    \centering
    \includegraphics{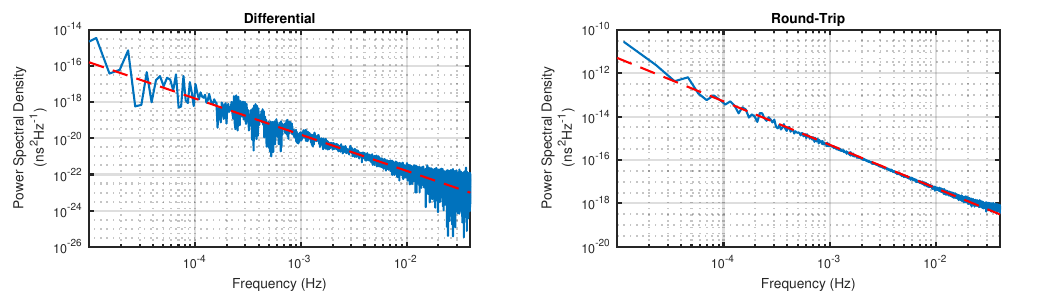}
     \caption{{Power spectral density of time-of-flight drift for (a) the Differential and (b) the Round-Trip Configurations. The red dashed lines are a reference 20~dB-per-decade expected for Brownian noise.}}
    \label{fig:time_psd}
\end{figure}
 {
\section{Power Spectral Densities}
In Figures~\ref{fig:phase_psd}--\ref{fig:time_psd} we show the power spectral densities of the phase and time-of-flight drifts, respectively. These plots also provide a 20-dB-per-decade reference line (red dashed), the expected profile for noise caused by a Brownian process. The good agreement between this model and our spectra indicates that our drifts are well-described by primarily Brownian processes.}

\section{Transmitter and Receiver Sequences}

\begin{figure*}[h!]
\centering
\includegraphics{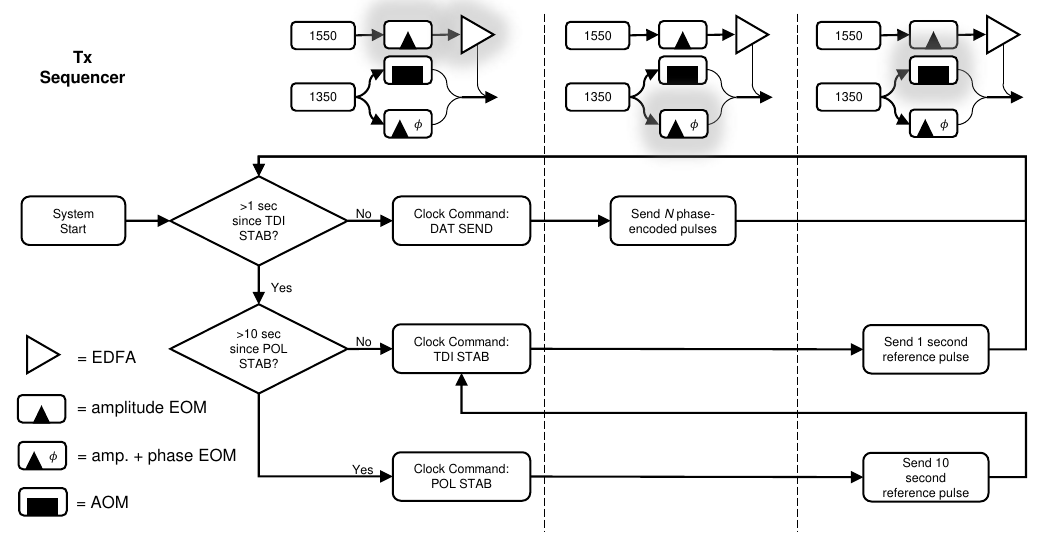}
\caption[Transmitter experimental sequence.]{\textbf{Transmitter experimental sequence.} Logical flowchart for the AWG sequencer used to control Alice's transmitter. Inset diagrams show a simplified version of the transmitter hardware, with glows indicating components which are being actively controlled during given portions of the sequence.}
\label{fig:sequence_tx}
\end{figure*}

\begin{figure*}
\centering
\includegraphics{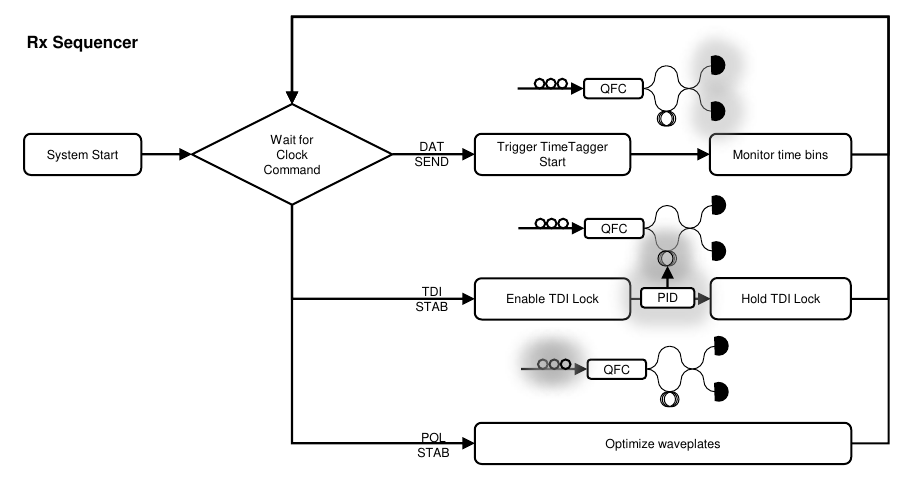}
\caption{\textbf{Receiver experimental sequence.} Logical flowchart for the AWG sequencer used to control Bob's receiver. Inset diagrams show a simplified version of the receiver hardware, with glows indicating components which are being actively controlled during given portions of the sequence.}
\label{fig:sequence_rx}
\end{figure*}

The experimental sequences at the transmitter (Tx) and receiver (Rx) are controlled each by a local arbitrary waveform generator (HD-AWG, Zurich Instruments). Figures~\ref{fig:sequence_tx} and Figures~\ref{fig:sequence_rx} depict the logic flowchart followed by each of these sequencers.

\bibliography{references}
\noindent [47] S. M. Joffre and T. Laurila, Standard deviations of wind speed and direction from observations over a smooth surface, Journal of Applied Meteorology and Climatology 27, 550 (1988).